\documentclass[12pt]{article}
\usepackage[a4paper, total={7in, 10in}]{geometry}
\usepackage[parfill]{parskip}
\usepackage{physics, tensor, float, subcaption}
\usepackage{graphicx}
\graphicspath{ {Plots/} }
\usepackage{jhep-mod}
\usepackage{bm}
\usepackage{soul}
\usepackage{amssymb,amsmath,amsthm}
\usepackage{mathrsfs}
\usepackage[utf8]{inputenc}
\usepackage{enumerate}
\usepackage{bigints}
\usepackage{xcolor}
\usepackage{appendix}
\usepackage{graphicx}
\usepackage{float}
\usepackage{tikz}
\usepackage{setspace}
\usepackage{cancel}
\usepackage{array}
\usepackage{tabulary}
\usepackage{doi}
\definecolor{purple}{rgb}{1,0,1}
\definecolor{lime}{HTML}{A6CE39} 

\definecolor{lime}{HTML}{A6CE39}
\newcommand{\orcidicon}{%
	\begin{tikzpicture}
	\draw[lime, fill=lime] (0,0) 
		circle [radius=0.16] 
		node[white] {{\fontfamily{qag}\selectfont \tiny ID}};
	\draw[white, fill=white] (-0.0625,0.095) 
		circle [radius=0.007];
	\end{tikzpicture}
	\hspace{-5mm}
}
\newcommand\orcidJosh{{\href{https://orcid.org/0000-0003-1200-7261}{\orcidicon}}}
\newcommand\orcidMatt{{\href{https://orcid.org/0000-0003-1088-6485}{\orcidicon}}}

\renewcommand{\O}{\mathcal{O}}

\newcommand{\be}{\begin{equation}}
\newcommand{\ee}{\end{equation}}

\newcommand{\n}{\nabla}
\begin{document}


\title{\vspace{-25pt}\huge{
Killing horizons and surface gravities for a well-behaved three-function generalisation of the Kerr spacetime 
}}


\author{
\Large
Joshua Baines\!\orcidJosh {\sf  and} Matt Visser\!\orcidMatt}
\affiliation{School of Mathematics and Statistics, Victoria University of Wellington, 
\\
\null\qquad PO Box 600, Wellington 6140, New Zealand.}
\emailAdd{joshua.baines@sms.vuw.ac.nz}
\emailAdd{matt.visser@sms.vuw.ac.nz}

\abstract{
\vspace{1em}

Thanks to the recent advent of the event horizon telescope (EHT), 
we now have the opportunity to test the physical ramifications of the 
strong-field near-horizon regime for astrophysical black holes. 
Herein, emphasizing the trade-off between tractability and generality, 
the authors discuss a particularly powerful three-function distortion of the Kerr spacetime, 
depending on three arbitrary functions of the radial coordinate $r$, 
which on the one hand can be fit to future observational data,
and on the other hand is sufficiently general so as to encompass an extremely wide class of theoretical models. 
In all of these spacetimes, both  the timelike Hamilton--Jacobi (geodesic) and massive Klein--Gordon (wave) equations separate, 
and the spacetime geometry is asymptotically Kerr; hence these spacetimes are well-suited 
to modelling real astrophysical black holes. The authors then prove the existence of Killing horizons 
for this entire class of spacetimes, and give tractable expressions for the angular velocities, 
areas, and surface gravities of these horizons. 
We emphasize the validity of rigidity results and zeroth laws for these horizons.

\bigskip
\noindent
{\sc Date:} 13 March 2023; 29 March 2023; 15 May 2023; \LaTeX-ed \today

\
Killing tensor; Carter constant; Klein--Gordon separability; 
Killing--Yano tensor; \\
Killing horizon; surface gravity; zeroth law.

\bigskip
\noindent{\sc PhySH:} 
Gravitation

\bigskip
Published as: Universe \textbf{9 \# 5} (2023) 223; \doi{10.3390/universe9050223}; \\arXiv:2303.07380 [gr-qc]
}


\maketitle
\def\tr{{\mathrm{tr}}}
\def\diag{{\mathrm{diag}}}
\def\cof{{\mathrm{cof}}}
\def\pdet{{\mathrm{pdet}}}
\parindent0pt
\parskip7pt
\section{Introduction}

In general relativity  the most astrophysically important exact vacuum solution to the Einstein equations is arguably the Kerr solution~\cite{Kerr,Kerr-Texas,Newman:1965,Boyer:1966,Carter:1968,Robinson:1975,Doran:1999,kerr-book-1,kerr-book-2,kerr-intro,Kerr:discovery,Teukolsky:2014,Adamo:2014,Ansatz,Darboux}. 
See also the textbook presentations in~\cite{Adler-Bazin-Schiffer,D'Inverno,Hartle,Carroll,Wald,Weinberg,Hobson,MTW}. 
This exact solution models a stationary, axially symmetric spacetime and hence models the vacuum, exterior spacetime of a rotating, axially symmetric black hole. 
For the non-rotating case, the spacetime is modelled by the Schwarzschild solution~\cite{Schwarzschild:1916} which,
due to Birkhoff's theorem~\cite{Birkhoff,Jebsen,Deser:2004, Ravndal:2005}, 
is the unique spherically symmetric vacuum solution to Einstein's equations.
However, Birkhoff's theorem does not hold for a (3+1) dimensional  axially symmetric vacuum spacetime \cite{kerr-intro, Skakala, PGLT1}. While the Kerr solution models the exterior spacetime of a rotating general relativistic black hole to a high degree of accuracy, it is not guaranteed that the Kerr solution will model exterior spacetime geometry for all astrophysical rotating compact objects (especially in the strong field regime).

\clearpage
Such issues have become increasingly pressing now that we have been able to partially resolve the near horizon structure of some astrophysical black holes with the EHT --- the event horizon telescope~\cite{EHT-1, EHT-4, EHT-5, EHT-SgA*-1, EHT-SgA*-6, Psaltis:2014, Broderick:2013, Cardoso:2016, Carballo-Rubio:2018a, Carballo-Rubio:2018b}.
We are now in a position to begin to test the applicability of the Kerr solution in the strong field regime of astrophysical black holes.

When performing such probes of the Kerr spacetime, it is extremely useful to have a tractable template to work with, one that is not too general nor too restrictive. 
These distorted-Kerr templates have typically been  studied at the level of phenomenological \emph{models}, where one is primarily guided by the key symmetries, and by requiring the ability to continuously connect one's distorted-Kerr spacetime back to the exact Kerr geometry. More recently there has been some effort devoted to studying the dynamics of such models, attempting to either find or manually construct a suitable Lagrangian. For instance, many regular black holes, originally proposed as purely phenomenological \emph{models}, now have dynamical interpretations in terms of nonlinear electrodynamics~\cite{Toshmatov:2017a,Toshmatov:2017b,Toshmatov:2018ccc,Rodrigues:2017, Yu:2019,Breton:2019,Canate:2022a,Canate:2022b,Bronnikov:2022,Kubiznak:2022,Franzin:2022}. Similarly, extending the results of~\cite{Boonserm:2015} to deal with energy condition violations, the phenomenologically inspired ``black bounce'' spacetimes now have a dynamical interpretation in terms of a combination of phantom scalars and nonlinear electrodynamics~\cite{Bronnikov:2021a,Bronnikov:2022ccc,Rodrigues:2023a}. See also~\cite{Huang:2019,Huang:2020,Bronnikov:2022eee}. Work on these issues is ongoing. 

Herein, we focus attention on a triply-infinite (3-function) class of stationary, axially symmetric spacetimes which can be tested against data of exterior spacetimes of real, astrophysical, rotating black holes. We prove the existence of Killing horizons in all of these spacetimes, and give compact and tractable expressions for the horizon area and surface gravity of these modified black holes, verifying that the zeroth law of black hole mechanics is satisfied. 
Ultimately, since we wish our candidate black hole spacetimes to model the gravitational fields of real astrophysical black holes, some observationally inspired physical constraints must be placed on these general black hole spacetimes. 
We shall impose:
\begin{itemize}
\item Hamilton--Jacobi timelike separability.
\item Klein--Gordon  separability.
\item Asymptotic flatness.
\end{itemize}
We shall \emph{not} impose:
\begin{itemize}
\item Dirac separability.
\end{itemize}
We shall explicitly demand that all of our distorted-Kerr templates contain the exact Kerr spacetime as a special case.
Stability analyses for these distorted-Kerr templates are somewhat subtle. One has to carefully distinguish the notion of the stability of test fields perturbing these spacetimes, from the question of the internal stability of the spacetime itself. Test-field stability, at the level of the Regge--Wheeler and Zerelli equations, has been addressed in references~\cite{Boonserm:2013,Saleh:2018,Villani:2021}.
Once one specifies an appropriate class of gravitational sources, full dynamical stability can also be addressed --- see for example~\cite{Toshmatov:2019c, Toshmatov:2018c, Nomura:2020, Franzin:2022}.

\subsection{Hamilton--Jacobi timelike separability}
\enlargethispage{40pt}
Firstly, we wish to impose that the Hamilton--Jacobi equation is timelike separable, since timelike separability of this equation implies the existence of a non-trivial Killing tensor, which then gives rise to a fourth constant of the motion, hence making the timelike geodesic equation integrable. We have good physical evidence to impose separability of the Hamilton--Jacobi equation. Separability of the Hamilton--Jacobi equation excludes the possibility for chaotic particle orbits, we wish to impose this since we certainly have observed long--lived accretion disks. These accretion disks would rapidly destabilise if we had chaotic particle orbits. As shown by Papadopoulos and Kokkotas \cite{Papadopoulos:2020,Papadopoulos:2018}, based on earlier work by Benenti and Francaviglia~\cite{Benenti:1979}, (see also~\cite{PGLT2}),  to guarantee the separability of the (timelike) Hamilton--Jacobi equation, for a general stationary axisymmetric spacetime we have to impose the condition that the inverse metric can be written, in $(r,\theta,\phi,t)$ coordinates, in the following form:
\begin{equation} \label{PK-10}
g^{ab}(r,\theta) = 
{1\over  A_1 + B_1}
\left[\begin{array}{cccc}
A_2 & 0 & 0 & 0\\ 
0 & B_2 & 0 & 0\\
0 & 0 & A_3+B_3 & \;A_4+B_4\\
0 & 0 & A_4+B_4 & \;A_5+B_5
\end{array}\right] .
\end{equation}
This contravariant metric  contains ten arbitrary single-variable functions, the five $A_i$ functions depend only on $r$, and the five $B_i$ functions depend only on $\theta$. 

The metric (\ref{PK-10}) has the following Killing tensor
\begin{equation}
K^{ab}(r,\theta) = 
{1\over  A_1 + B_1}
\left[\begin{array}{cccc}
A_2B_1 & 0 & 0 & 0\\
0 & -B_2A_1 & 0 & 0\\
0 & 0 & B_1A_3-A_1B_3 &\; B_1A_4-A_1B_4\\
0 & 0 & B_1A_4-A_1B_4 & \;B_1A_5-A_1B_5
\end{array}\right] .
\end{equation}
Independently, Carson and Yagi have also generated a 10-function contravariant metric that separates the Hamilton--Jacobi equation \cite{Carson:2020}. Their metric, ultimately, is a specific implementation of the 10-function metric proposed by Papadopoulos and Kokkotas, equation \eqref{PK-10}. 
We point out that Hamilton--Jacobi separability is known to have direct physical implications for photon orbits~\cite{Pappas:2018,Pappas:2019}.

Working from another direction, Johannsen was able to show that a certain 8-function metric also separates the Hamilton--Jacobi equation \cite{Johannsen:2015a}. This 8-function metric is a restriction of the 10-function 
Papadopoulos--Kokkotas metric. Mathematically the restriction can be phrased in terms of demanding that the metric determinant be a perfect square. Physically this corresponds to demanding the existence of a relatively tractable tetrad.
Johannsen then imposed further physical constraints~\cite{Johannsen:2015b}, by analysing the metric the context of the PPN framework and by examining the asymptotic behaviour of the metric. By imposing these constraints, Johannsen was able to produce a 4-function metric satisfying both Hamilton--Jacobi integrability and asymptotic flatness~\cite{Johannsen:2015b}. 
See also references~\cite{Lima:2020a,Shaikh:2021} for an approach based on the Newman--Janis ansatz.

\subsection{Klein--Gordon separability}

Secondly, we wish to additionally impose the separability of the wave equation (the Klein--Gordon equation). 
It can be shown that this condition is satisfied if the commutator 
\be
\left[K,R\right]^a{}_b\equiv K^a{}_c R^c{}_b-R^a{}_c K^c{}_b
\ee
is a conserved $T^1_1$ tensor~\cite{Commutator:2002,Giorgi:2021,Baines:2021,Franzin:2021}, see also~\cite{Carter:1968b,Teukolsky:1972,Kalnins:1980}.
That is, a necessary and sufficient condition for Klein--Gordon separability is 
\be \label{Wave_eqn_con}
\n_a\left[K,R\right]^a{}_b\equiv\n_a\left(K^a{}_c R^c{}_b-R^a{}_c K^c{}_b\right)=0 ,
\ee
where $K_{ab}$ is the Killing tensor of the spacetime and $R_{ab}$ is the Ricci tensor. 
This condition can be manipulated into a constraint on the $A_i$ functions, see \cite{Papadopoulos:2020}, so that only three of the five $A_i$ are functionally independent. See also~\cite{KSZ-3,Konoplya:2016jvv}.

\enlargethispage{40pt}
The physical reason we demand separability of the wave equation is to ensure the existence of quasi--normal modes which arise when considering ring--down effects which occur after black hole merger events \cite{QNM-1, QNM-2}. 
See also reference~\cite{Chen:2019} for an approach based on the Newman--Janis ansatz.

\subsection{Asymptotic flatness}

Thirdly, we will explicitly demand asymptotic flatness --- 
in particular that all of the $B_i$ functions are exactly those of the Kerr spacetime. If this were not the case, the angular dependence of the spacetime time would not asymptotically approach Kerr for large $r$. 
Even though the techniques used, and the mathematical formulations employed, are significantly different, there is nevertheless widespread agreement that the combination of Hamilton--Jacobi separability, Klein--Gordon separability, and asymptotic flatness ultimately leads to a distorted Kerr framework with exactly three free functions~\cite{Papadopoulos:2020,KSZ-3,Konoplya:2016jvv}. 

We shall now present a computationally efficient version of the 3-function distorted Kerr metric, with a particular emphasis on verifying the existence of Killing horizons, verifying the existence of a rigidity result, explicitly calculating the surface gravity, and verifying that this entire class of 3-function distorted Kerr spacetimes satisfies the zeroth law of black hole mechanics.

\subsection{Dirac separability --- a bridge too far}

But let us first say a word or two regarding a condition we will \emph{not} be imposing --- Dirac separability~\cite{Iyer:1985,Rudiger:1984,McLenaghan}.  While separability of the Dirac equation would be nice to have, it is nowhere near as critical to observational astrophysics as Hamilton--Jacobi (geodesic) separability or Klein--Gordon (wave) separability. Because of the Pauli exclusion principle, Fermi fields do not have a macroscopic classical limit --- while one might see individual fermions emitted by a black hole, there is no meaningful manner in which coherent Fermion emission might occur. Certainly the Dirac equation partially separates, due to the time-like and azimuthal Killing vectors, but there is no pressing physical need to demand complete Dirac separability in the remaining $(r,\theta)$ plane.
In contrast, complete separability of the geodesic and wave equations has direct observational significance. 


\section{Three-function generalisation of the Kerr spacetime}
We now start our analysis of the 3-function distorted Kerr spacetime by considering the following co-tetrad:
\be
\begin{split}
e^0=\sqrt{\frac{\Delta_0}{\Sigma_0}}\;(dt-a\sin^2(\theta)d\phi); \quad e^1=\sqrt{\frac{\Sigma_0}{\Delta_0}}\;dr;\\
e^2=\sqrt{\Sigma_0}\;d\theta; \quad e^3=\frac{\sin(\theta)}{\sqrt{\Sigma_0}}\;(-a\,dt+(r^2+a^2)d\phi) .
\end{split}
\ee
Here we set $\Delta_0=r^2+a^2$ and $\Sigma_0=r^2+a^2\cos^2(\theta)$. The corresponding metric tensor is $g_{ab}=e^A{}_a\;e^B{}_b\;\eta_{AB}$, which is just Minkowski spacetime in disguise. (In oblate spheroidal coordinates, which reduce to spherical polar coordinates as $a\to0$.) 

However, we if now introduce three arbitrary functions of $r$, namely $\Delta(r)$, $\Phi(r)$, and $\Xi(r)$, and furthermore set $\Sigma(r,\theta)=\Xi(r)^2+a^2\cos^2(\theta)$, we can consider a modified (distorted) co-tetrad
\be \label{3f-cotetrad}
\begin{split}
e^0=\exp(-\Phi(r))\;\sqrt{\frac{\Delta(r)}{\Sigma(r,\theta)}}\;(dt-a\sin^2(\theta)d\phi); \quad 
e^1=\sqrt{\frac{\Sigma(r,\theta)}{\Delta(r)}}\; dr;\\
e^2=\sqrt{\Sigma(r,\theta)}\; d\theta; \quad 
e^3=\frac{\sin(\theta)}{\sqrt{\Sigma(r,\theta)}}\; (-a\,dt+(\Xi(r)^2+a^2)d\phi) .
\end{split}
\ee
The resulting line element is now
\begin{equation} \label{3f_kerr_metric}
\begin{split}
ds^2= & -\frac{\Delta(r)\exp(-2\Phi(r))-a^2\sin^2(\theta)}{\Xi(r)^2+a^2\cos^2(\theta)}\; dt^2
+\frac{\Xi(r)^2+a^2\cos^2(\theta)}{\Delta(r)}\; dr^2\\
&+(\Xi(r)^2+a^2\cos^2(\theta))\;d\theta^2
-2\;\frac{a \sin^2(\theta) \;(\Xi(r)^2-\Delta(r)\exp(-2\Phi(r))+a^2)}{\Xi(r)^2+a^2\cos^2(\theta)}\;dtd\phi\\
&+\frac{ \left((\Xi(r)^2+a^2)^2-\exp(-2\Phi(r))\Delta(r)a^2\sin^2(\theta)\right)\sin^2(\theta)}{\Xi(r)^2+a^2\cos^2(\theta)}\; d\phi^2 .
\end{split}
\end{equation}
Note that when we set $\Delta(r)=r^2-2mr+a^2$, $\Phi(r)=0$, and $\Xi(r)=r$, we recover the Kerr spacetime written in Boyer-Lindquist coordinates. In order to satisfy the condition that this spacetime is asymptotically Kerr we impose the following conditions on our three arbitrary functions of $r$ 
\be
\Delta(r)\sim r^2; \qquad \Phi=o\left(1\right); \qquad \Xi(r)\sim r.
\ee

It is perhaps worthwhile to point out that the zero rotation $a\to0$ limit is particularly pleasant. For the cotetrad
\be \label{3f-cotetrad-zero}
\begin{split}
e^0\to\exp(-\Phi(r))\;{\frac{\sqrt{\Delta(r)}}{\Xi(r)}}\;dt; \qquad 
e^1\to\frac{\Xi(r)}{\sqrt{\Delta(r)}}\; dr;\\
e^2\to\Xi(r)\; d\theta; \qquad 
e^3\to \Xi(r) \sin(\theta)\; d\phi,
\end{split}
\ee
while for the line element
\be
ds^2 \to 
\exp(-2\Phi(r))\;{\frac{{\Delta(r)}}{\Xi(r)^2}}\;dt^2 + {\Xi(r)^2\over\Delta(r)} \; dr^2 
+ \Xi(r)^2 [d\theta^2 +\sin^2(\theta) \; d\phi^2]. 
\ee

This three-function class of spacetimes is generically no longer Ricci flat, however it covers a vast class of physically interesting spacetimes, as outlined in Table~\ref{T:1} below.

\begin{table}[h!]
\caption{Various spacetimes that fall into the 3-function class discussed herein.\vphantom{\Big |}}
\hspace{-30pt}
\begin{tabular}{|>{\centering\arraybackslash}m{5.5cm}|>{\centering\arraybackslash}m{4.5cm} >{\centering\arraybackslash}m{1.6cm} >{\centering\arraybackslash}m{1.8cm} >{\centering\arraybackslash}m{1.6cm}|}
\hline
\hline
Spacetime & $\Delta(r)$ & $\Phi(r)$ & $\Xi(r)$ & $a$ \\ 
 \hline
 \hline
 Minkowski & $r^2+a^2$ & 0 & $r$ & arbitrary\\ 
 \hline
 Kerr & $r^2+a^2-2mr$ & 0 & $r$ & non-zero \\
 Kerr--Newman & $r^2+a^2-2mr+Q^2$ & 0 & $r$ & non-zero \\
 Eye of storm & $r^2-2e^{-l/r} mr+a^2$ & 0 & $r$ & non-zero\\
 Carter 1-function off-shell & arbitrary & 0 & $r$ & non-zero \\
 {Sugra STU ``balanced"} & {$r^2-2mr+a^2$}& 0  & {$\sqrt{r^2-2mr}$}& non-zero\\
 Kerr black-bounce & $r^2+l^2-2m\sqrt{r^2+l^2}+a^2$ & 0 & $\sqrt{r^2+l^2}$ & non-zero \\
 \hline
 Schwarzschild & $r^2-2mr$ & 0 & $r$ & 0 \\
 Reissner--Nordstr\"om & $r^2-2mr+Q^2$ & 0 & $r$ & 0 \\
 Kiselev & arbitrary & 0 & $r$ & 0\\
 Static spherical symmetry  & arbitrary & arbitrary & arbitrary & 0 \\
 Morris--Thorne wormhole & $r^2+l^2$ & 0 & $\sqrt{r^2+l^2}$ & 0 \\
 Simpson--Visser black bounce & $r^2+l^2 -2m\sqrt{r^2+l^2}$ & 0 & $\sqrt{r^2+l^2}$ & 0 \\
 exponential wormhole & $r^2$ & $0$ & $r \, e^{m/r}$ & 0\\
\hline
\hline
\end{tabular}
\label{T:1}
\end{table}

\bigskip
\bigskip
Note that the Kerr black-bounce  geometry~\cite{Mazza:2021,Islam:2021} is a rotating version of the Simpson--Visser black bounce~\cite{Simpson:2018}, which in turn generalizes the Morris--Thorne traversable wormhole~\cite{Morris-Thorne,MTY,Lorentzian-wormholes}.
The supergravity inspired STU ``balanced'' (equal charge) black holes of reference~\cite{STU} are also a subset of the 3-function class considered herein. 
However the generic ``unbalanced'' STU black holes of reference~\cite{STU}, which have null separable geodesics, but not timelike separable geodesics,  do not fall into our 3-function class. (See also reference~\cite{Keeler:2012}.)

The Carter 1-function off-shell spacetime (more on this below) is an asymptotically Kerr restriction of Carter's canonical 2-function off-shell spacetime~\cite{Carter:1968b, Frolov_et_al,Frolov:2006}. In the other direction the Carter 1-function off-shell spacetime can further be restricted to the ``eye of storm'' spacetime~\cite{Eye:1,Eye:2,Ghosh:2014} which is particularly ``close to Kerr''. 

In counterpoint, Kiselev spacetimes~\cite{Kiselev:2002,Visser:Kiselev,Rodrigues:2022kiselev,Semiz:2020}  generalize the Schwarzschild and Reissner--Nordstr\"om spacetimes, whereas  the static spherically symmetric spacetimes with arbitrary $\Xi(r)$ are sufficiently general to enable one to deal with static spherically symmetric traversable wormholes~\cite{Morris-Thorne,MTY,Lorentzian-wormholes, Examples, Surgery, Lemos:2003, Lobo:2005, Lobo:2005bb, Teo:1998, Bueno:2017, Simpson:2019b, Berry:2020, Lobo:2020, Bambi:2021, Maeda:2021,Clement:2022, exponential}
(Note that restricting $\Xi(r)\to r$ is more appropriate for stars and planets.)

Indeed we note that the general 3-function spacetime (\ref{3f_kerr_metric}) is a specific example of the Hamilton--Jacobi separable spacetime \eqref{PK-10}, where we can identify 
\begin{gather}
A_1=\Xi(r)^2; \qquad A_2=\Delta(r); \qquad A_3=-\frac{a^2\exp(2\Phi(r))}{\Delta(r)}; \\
A_4=-\frac{a\exp(2\Phi(r))(\Xi(r)^2+a^2)}{\Delta(r)}; \quad A_5=-\frac{\exp(2\Phi(r))(\Xi(r)^2+a^2)^2}{\Delta(r)} ,
\end{gather}
and 
\be
B_1=a^2\cos^2(\theta); \quad B_2=1; \quad B_3=\frac{1}{\sin^2(\theta)}; \quad B_4=a; \quad B_5=a^2\sin^2(\theta) .
\ee
We emphasize that the $B_i$ functions are exactly those of Kerr, this is intentional. 
In particular note that $B_3 B_5 = B_4^2$. 
As advertised, only three of the $A_i$ functions are functionally independent; in  particular $A_3 A_5 = A_4^2$.

\enlargethispage{20pt}
We now calculate the non-trivial Killing tensor in $(t, r, \theta, \phi)$ coordinates
\begin{equation} \label{KT}
K^{ab}(r,\theta) = 
\left[\begin{array}{cccc}
K^{tt}&0&0&K^{t\phi}\\ 
0 &  \frac{\Delta(r)a^2\cos^2(\theta)}{\Xi(r)^2+a^2\cos^2(\theta)} &0&0 \\
0 & 0 & -\frac{\Xi(r)^2}{\Xi(r)^2+a^2\cos^2(\theta)}&0\\
K^{t\phi}&0&0&K^{\phi\phi}
\end{array}\right] ,
\end{equation}
where
\be
\begin{split}
\label{KT2}
K^{tt}= & -\frac{a^2\left(\exp(2\Phi(r))(\Xi(r)^2+a^2)^2\cos^2(\theta)+\Xi(r)^2\Delta(r)\sin^2(\theta)\right)}{\Delta(r)(\Xi(r)^2+a^2\cos^2(\theta))} ;\\
K^{t\phi}= & -\frac{a\left(a^2\exp(2\Phi(r))(\Xi(r)^2+a^2)\cos^2(\theta)+\Xi(r)^2\Delta(r)\right)}{\Delta(r)(\Xi(r)^2+a^2\cos^2(\theta))} ;\\
K^{\phi\phi}= &\; -\frac{a^4\exp(2\Phi(r))\cos^2(\theta)\sin^2(\theta)+\Xi(r)^2\Delta(r)}{\Delta(r)\sin^2(\theta)(\Xi(r)^2+a^2\cos^2(\theta))} .
\end{split}
\ee
It is then not hard to check that $\n_{(c}K_{ab)}=0$. Hence this object is indeed a Killing tensor, and its existence ensures that the Hamilton--Jacobi equation separates.

\noindent To check that the wave equation (Klein--Gordon equation) separates, we need to check that equation \eqref{Wave_eqn_con} is satisfied. While the commutator $[K,R]^{ab}$ is non-zero, it is relatively simple.
Axi-symmetry (plus very mild circularity conditions on the Killing vectors) is enough to ensure the in $(t,r,\theta,\phi)$ coordinates both the metric and inverse metric are symmetric and of the form~\cite{Wald}
\begin{equation}
\left[\begin{array}{cccc}
\;*\;&\;0\;\;&\;0\;&\;*\\ 
0 & * &0&0 \\
0 & 0 & *&0\\
*&0&0&*
\end{array}\right] .
\end{equation}
Hamilton-Jacobi separability then guarantees that both $K^{ab}$ and $K_{ab}$ are of this same form.
For any spacetime in our 3-function distortion of Kerr both $R^{ab}$ and $R_{ab}$ are also of this same form.
(This result for $R^{ab}$ and $R_{ab}$ does not hold if one merely assumes Hamilton--Jacobi separability.) 
Under the stated conditions this is enough to imply that the commutator is then of the even more restrictive form
\begin{equation}
[K,R]^{ab} = 
\left[\begin{array}{cccc}
0&\;0\;&\;0\;&[K,R]^{t\phi}\\ 
0&0&0&0 \\
0&0&0&0\\
-[K,R]^{t\phi}&0&0&0
\end{array}\right].
\end{equation}
But then
\be
\begin{split}
\nabla_a 
[K,R]^{ab} 
&= {1\over\sqrt{-g}} \partial_a (\sqrt{-g}[K,R]^{ab} )\\
& = 
{1\over\sqrt{-g}} 
\Big(- \partial_\phi \{\sqrt{-g} [K,R]^{t\phi} \},
0,0, \partial_t \{\sqrt{-g} [K,R]^{t\phi} \}
\Big) = 0.
\end{split}
\ee
So the Klein--Gordon (wave) equation is indeed separable as advertised. 

For completeness we note that explicit calculation yields
\be
[K,R]^{t\phi}=-\frac{a\left[\exp(\Phi(r))\;{d\over dr} \left(\exp(\Phi(r))\;\Xi(r)\;\frac{d\Xi(r)}{dr}\right)-1\right]}{\Xi(r)^2+a^2\cos^2(\theta)} .
\ee
Thus the commutator is  independent of $\Delta(r)$ and vanishes if both $\Phi(r)\to0$ and $\Xi(r)\to \pm r$; an observation that will subsequently be of use when discussing the possibility of having Killing--Yano tensors. 

\section{Killing horizons for the 3-function generalization of Kerr}
Let us now focus on near-horizon and on-horizon physics~\cite{Hawking:2014,observability,Medved:2004}.

A plausible tentative location for the horizons of this spacetime is that they occur when the $g_{rr}$ component of the metric becomes singular, these locations are specified by
\be \label{rH}
\{r_{H_i}:\Delta(r_{H_i})=0\} .
\ee
Since on these putative horizons
\be
g^{ab} \, \partial_a r \, \partial_b r = g^{rr} = {\Delta(r)\over\Sigma(r,\theta)} \to 0
\ee
we see that these putative horizons are certainly null hypersurfaces.

For these horizons to be event horizons, they must at a minimum be invariant under both time translations and axial rotations. This is equivalent to the condition that geodesics on the horizon moving in either the $t$ or $\phi$ direction should stay on the horizon. More generally, geodesics on the horizon moving in the $t, \theta$ or $\phi$ direction should stay on the horizon. That is
\be
\left. \frac{d^2r}{d\lambda^2}\right|_{r=r_{H_i}}=\left.-\Gamma^{r}{}_{ij}\frac{dx^i}{d\lambda}\frac{dx^j}{d\lambda}\right|_{r=r_{H_i}}=0 ,
\ee
where $i=\{t,\theta,\phi\}$ and $j=\{t,\theta,\phi\}$. That is, we wish that 
$\Gamma^r{}_{ij}$ 
vanishes on the horizon. 
But
\be
\Gamma^r{}_{ij}= g^{rr} \left( g_{r(i,j)} -\textstyle{1\over2} g_{ij,r}\right) = -\textstyle{1\over2} g^{rr} g_{ij,r}\propto \Delta(r).
\ee
So certainly $\Gamma^r{}_{ij}\to 0$ on the putative horizons.  
In fact a brief explicit  calculation yields
\be
\Gamma^r{}_{tt}\propto \Delta(r);\quad \Gamma^r{}_{t\phi}\propto \Delta(r);\quad 
\Gamma^r{}_{\phi\phi}\propto \Delta(r);\quad \Gamma^r{}_{\theta\theta}\propto \Delta(r);\;
\ee
and 
\be
\Gamma^r{}_{t\theta}=0;\qquad \Gamma^r{}_{\theta\phi}=0 .
\ee
So these putative horizons have the same physical effect as event horizons. \\

To prove that these horizons are indeed Killing horizons, we need to prove the existence of a Killing vector whose norm is null when evaluated on the horizons. There exist two Killing vectors in this spacetime. One associated with time translation symmetry
\be
\xi^a=(1,0,0,0)^a ,
\ee
and one associated with axial symmetry 
\be
\psi^a=(0,0,0,1)^a .
\ee
One can construct a more general Killing vector by taking a linear combination of the two
\be
\label{E:K-generic}
K^a=(1,0,0,\Omega) ,
\ee
where $\Omega$ is a constant. We wish to find a collection of constants $\Omega_{{H_i}}$ such that the norm of $[K_{H_i}]^a=(1,0,0,\Omega_{{H_i}})$ vanishes on the relevant horizon $r_{H_i}$. That is we wish to impose 
\be
\left.g_{ab}K^aK^b\right|_{r=r_{H_i}}=\left.g_{tt}+2\Omega g_{t\phi}+\Omega^2g_{\phi\phi}\right|_{r=r_{H_i}}=0 .
\ee
A brief calculation yields
\be
g_{ab}K^aK^b=\frac{\left[(\Xi(r)^2+a^2)\Omega-a\right]^2\sin^2(\theta)}{\Xi(r)^2+a^2\cos^2(\theta)}-\Delta(r)\frac{\exp(-2\Phi(r))[1-a\Omega\sin^2(\theta)]^2}{\Xi(r)^2+a^2\cos^2(\theta)} ,
\ee
so that on the horizons
\be
\left.g_{ab}K^aK^b\right|_{r=r_{H_i}}=\frac{\left[(\Xi(r_{H_i})^2+a^2)\Omega-a\right]^2\sin^2(\theta)}{\Xi(r_{H_i})^2+a^2\cos^2(\theta)} .
\ee
Hence, in order for the horizons to be Killing we must demand that
\be
\Omega_{{H_i}}=\frac{a}{\Xi(r_{H_i})^2+a^2} .
\ee
Note that these constants are independent of $\theta$, hence these horizons rotate as if they are a rigid body, as we should expect. This rigidity result would in standard general relativity subject to suitable energy conditions be a rigidity \emph{theorem}, here it is instead a rigidity \emph{observation} based on our model for distorted Kerr.

\enlargethispage{30pt}
So we have shown that for these 3-function models there exist distinct Killing vectors $(K_{H_i})^a$ whose norm vanishes on the various horizons; thus the horizons specified by the condition in equation \eqref{rH} are indeed Killing horizons.
This assertion is by no means vacuous, there are many other stationary axisymmetric models which can be constructed for which the horizons are not necessarily Killing horizons~\cite{non-Killing,LRR,Acoustic,Fischetti:2016,Li:2021}.

\noindent The area of surfaces of constant $r$ in this spacetime are given by
\be
\begin{split}
S(r) & =  2\pi\int^{\pi}_0 \sqrt{g_{\theta\theta}\;g_{\phi\phi}}\;\sin(\theta)\, d\theta\\
& = 2\pi\int^{\pi}_0 \sqrt{(\Xi(r)^2+a^2)^2+\exp(-\Phi(r))\Delta(r)a^2\sin^2(\theta)}\;\sin(\theta)\, d\theta .
\end{split}
\ee
In general this integral is somewhat messy. However, at the horizons this equation simplifies greatly.  The areas of the horizons are explicitly given by
\be
S(r_{H_i})=2\pi\;(\Xi(r_{H_i})^2+a^2)\int^{\pi}_0\sin(\theta)\; d\theta=4\pi\;(\Xi(r_{H_i})^2+a^2) .
\ee

\section{Surface gravities for these Killing horizons}
The surface gravity for any Killing horizon is given by~\cite{Wald} 
\be
|\kappa|=\left.\sqrt{-\frac{1}{2}(\n_aK_b)(\n^aK^b)}\right|_{H}.
\ee
Here $K^a$ is the Killing vector whose norm vanishes on the horizon. 
We have already given an expression for such a Killing vector above in equation (\ref{E:K-generic}).
In view of Killing's equation we can also write this as 
\be
|\kappa|=\left.\sqrt{\frac{1}{2}(\n_aK_b)(\n^bK^a)}\right|_{H} = \left.\sqrt{\frac{1}{2}(\n_aK^b)(\n_bK^a)}\right|_{H}.
\ee

\enlargethispage{20pt}
Explicit calculation then gives the remarkably simple, and quite straightforward, expression
\be
\kappa_{H_i}=\frac{\exp(-\Phi(r_{H_i}))\;\Delta'(r_{H_i})}{2(\Xi(r_{H_i})^2+a^2)} .
\ee
Here the prime denotes differentiation with respect to $r$, and the sign has been chosen such that the surface gravity of the outermost horizon is positive.  

This expression is independent of $\theta$ and hence the zeroth law is satisfied. Furthermore, this expression can easily be checked against all of the well-known specific examples in the 3-function class of spacetimes we are considering.

Note that for the product of surface gravity and horizon area we have 
\be
\kappa_{H_i} \;S (r_{H_i})=2\pi\exp(-\Phi(r_{H_i}))\;\Delta'(r_{H_i}) .
\ee
It is convenient is some cases to write $\Delta(r)=r^2-2r\,m(r)+a^2$, then we have
\be
\kappa_{H_i}=\frac{\exp(-\Phi(r_{H_i}))\left[r_{H_i}{}^2(1-2m'(r_{H_i}))-a^2\right]}{2r_{H_i}(\Xi(r_{H_i})^2+a^2)} .
\ee
The zero-rotation limit, $a\to0$, then simplifies to
\be
\kappa_{H_i}\to \frac{\exp(-\Phi(r_{H_i})) \,r_{H_i}\,(1-2m'(r_{H_i}))}{2\,\Xi(r_{H_i})^2} .
\ee

\section{Two-function generalisation of the Kerr spacetime}

When one is not analysing physics on the throat (or anti-throat) of a wormhole, the areas of surfaces of constant $r$ must be a monotone function of $r$. 
(This is essentially the definition of a wormhole throat in the current stationary context.) 
Without further loss of generality, we can then set $\Xi(r)\to r$ and our metric becomes
\begin{equation} \label{2f_kerr_metric}
\begin{split}
ds^2= & -\frac{\Delta(r)\exp(-2\Phi(r))-a^2\sin^2(\theta)}{r^2+a^2\cos^2(\theta)}dt^2
+\frac{r^2+a^2\cos^2(\theta)}{\Delta(r)}dr^2\\
&+(r^2+a^2\cos^2(\theta))d\theta^2
+\frac{((r^2+a^2)^2-\exp(-2\Phi(r))\Delta(r)a^2\sin^2(\theta))\sin^2(\theta)}{r^2+a^2\cos^2(\theta)}d\phi^2\\
&-\frac{2a(r^2-\Delta(r)\exp(-2\Phi(r))+a^2)\sin^2(\theta)}{r^2+a^2\cos^2(\theta)}dtd\phi .
\end{split}
\end{equation}
\enlargethispage{20pt}
Our horizons are still given by the condition given in eqn \eqref{rH}, and they are still Killing horizons but the constant which makes the norm of $K^a=(1,0,0,\Omega)$ null on the horizons is now given by
\be
\Omega_{r_{H_i}}=\frac{a}{r_{H_i}{}^2+a^2} .
\ee
The area of surfaces of constant $r$ are now given by
\be
S(r) = 2\pi\int^{\pi}_0 \sqrt{(r^2+a^2)^2+\exp(-\Phi(r))\Delta(r)a^2\sin^2(\theta)}\;\sin(\theta)\, d\theta ,
\ee
and when evaluated on the horizons give
\be
S(r_{H_i})=2\pi\left(r_{H_i}{}^2+a^2\right)\int^{\pi}_0\sin(\theta)\, d\theta=4\pi\left(r_{H_i}{}^2+a^2\right) .
\ee
Our expression for the surface gravity is now given by
\be
\kappa_{H_i}=\frac{\exp(-\Phi(r_{H_i}))\Delta'(r_{H_i})}{2\left(r_{H_i}{}^2+a^2\right)} ,
\ee
where the prime again denotes differentiation with respect to $r$. The zeroth law is still satisfied. Note that we now have 
\be
\kappa_{H_i} \;S(r_{H_i})=2\pi\exp(-\Phi(r_{H_i}))\Delta'(r_{H_i}) .
\ee
It is convenient is some cases to write $\Delta(r)=r^2+a^2-2rm(r)$, then we have
\be
\kappa_{H_i}=\frac{\exp(-\Phi(r_{H_i}))\left[r_{H_i}{}^2(1-2m'(r_{H_i}))-a^2\right]}{2r_{H_i}\left(r_{H_i}{}^2+a^2\right)} .
\ee
If we let $a\rightarrow 0$ then $r_{H_i}\rightarrow 2m(r_{H_i})$ and we have
\be
\kappa_{H_i}=\exp(-\Phi(r_{H_i}))\;\frac{1-2m'(r_{H_i})}{2r_{H_i}} .
\ee
This is fully in agreement with known results in static spherical symmetry~\cite{DBH1}. 

\section{Would be Killing--Yano tensors}

A Killing--Yano tensor is an antisymmetric tensor $Y_{ab}$ that satisfies
\be
\n_{(c}Y_{a)b}=0 .
\ee
This implies that a Killing--Yano tensor provides  a `square root' of a Killing tensor, in the sense that
\be
Y_a{}^c\; Y_c{}^b=K_a{}^b
\ee
is a Killing tensor. 
The existence of a non-zero Killing--Yano tensor in a spacetime is an extremely non-trivial constraint and has significant physical implications~\cite{Frolov_et_al,Frolov:2006}. It can furthermore be shown that any Killing--Yano tensor commutes with the Ricci tensor~\cite{Lindstrom:2021,Lindstrom:2022a,Lindstrom:2022b}: 
\be
[Y,R]^a{}_b=Y^a{}_cR^c{}_b-R^a{}_cY^c{}_b=0 .
\ee
Thus the Killing tensor constructed from a Killing--Yano tensor $Y_{ab}$ will automatically satisfy $[K,R]^{ab}=0$, hence equation \eqref{Wave_eqn_con} is satisfied, and therefore the wave equation will be separable. 
Conversely if $[K,R]^{ab}\neq0$ then we cannot have $[Y,R]^{ab}=0$, and so there cannot be a Killing--Yano tensor. 

For our 3-function distortion of Kerr we start by first finding a `square root' of the Killing tensor, equation \eqref{KT}. Using our tetrad defined in equation \eqref{3f-cotetrad}, we can re-write the Killing tensor in a tetrad basis where it takes on a particularly simple (diagonal) form:
\begin{equation} \label{KT-tet}
K^{AB}(r,\theta) = 
\left[\begin{array}{cccc}
-a^2\cos^2(\theta)&0&0&0\\ 
0 & a^2\cos^2(\theta) &0&0 \\
0 & 0 & -\Xi(r)^2&0\\
0&0&0&-\Xi(r)^2
\end{array}\right] .
\end{equation}
Here indices in capital letters denote tetrad indices. We can then make an anzatz for our potential Killing--Yano tensor; in fact there are two distinct potential Killing--Yano tensors
\begin{equation} 
(Y_1)^{AB} = 
\left[\begin{array}{cccc}
0&a\cos(\theta)&0&0\\ 
-a\cos(\theta) &0&0&0 \\
0 & 0 & 0 & \Xi(r)\\
0&0&-\Xi(r)&0
\end{array}\right] ,
\end{equation}
and
\begin{equation} 
(Y_2)^{AB} = 
\left[\begin{array}{cccc}
0&-a\cos(\theta)&0&0\\ 
a\cos(\theta) &0&0&0 \\
0 & 0 & 0 & \Xi(r)\\
0&0&-\Xi(r)&0
\end{array}\right] .
\end{equation}
One can easily check that both $(Y_1)^{AC}(Y_1)_C{}^B=K^{AB}$ and 
$(Y_2)^{AC}(Y_2)_C{}^B=K^{AB}$, so these are two distinct ``square roots'' 
of the Killing tensor. In our coordinate basis we can write
\begin{equation} 
(Y_1)_{ab} = 
\left[\begin{array}{cccc}
0&Y_{tr}&-a\Xi(r)&0\\ 
-Y_{tr}&0&Y_{r\theta}&0 \\
a\Xi(r) & 0 & 0 & \Xi(r)(\Xi(r)^2+a^2)\\
0&-Y_{r\theta}&-\Xi(r)(\Xi(r)^2+a^2)&0
\end{array}\right] ,
\end{equation}
and
\begin{equation} 
(Y_2)_{ab} = 
\left[\begin{array}{cccc}
0&-Y_{tr}&-a\Xi(r)&0\\ 
Y_{tr}&0&-Y_{r\theta}&0 \\
a\Xi(r) & 0 & 0 & \Xi(r)(\Xi(r)^2+a^2)\\
0&Y_{r\theta}&-\Xi(r)(\Xi(r)^2+a^2)&0
\end{array}\right] ,
\end{equation}
where $Y_{tr}=a\exp(-\Phi(r))\cos(\theta)$ and $Y_{r\theta}=-a^2\exp(-\Phi(r))\cos(\theta)\sin^2(\theta)$. 

\enlargethispage{20pt}
Unfortunately, these two objects are not (in general) Killing--Yano tensors since one can check that (in general) $\n_{(c}(Y_1)_{a)b}\neq 0$ and $\n_{(c}(Y_2)_{a)b}\neq 0$. These objects are merely ``would be''  Killing--Yano tensors. 
However, there are two special sub-cases of interest:
\begin{itemize}
\item 
When $\Phi(r)=0$ and $\Xi(r)=-r$, as per our previous discussion $[K,R]^{ab}\to0$, and furthermore one has $\n_{(c}(Y_1)_{a)b}=0$, so $(Y_1)_{ab}$ then becomes a true Killing--Yano tensor. 
\item Similarly,  when $\Phi(r)=0$ and $\Xi(r)=r$, as per our previous discussion we also have $[K,R]^{ab}\to0$, and furthermore $\n_{(c}(Y_2)_{a)b}=0$, so $(Y_2)_{ab}$ then becomes a true Killing--Yano tensor. 
\end{itemize}
But in general, there does not exist a Killing--Yano tensor for the full 3-function generalisation of the Kerr spacetime given by equation \eqref{3f_kerr_metric}.

\section{Carter canonical off-shell metric}
\subsection{2-function version}

From another direction, considerable work has gone into understanding the geometry of the Carter canonical 2-function off-shell metric~\cite{Carter:1968,Carter:1968b, Frolov_et_al, Frolov:2006}. Carter's 2-function metric overlaps with the 3-function metric considered herein, but is not a subset thereof. There is instead a 1-function overlap.

Carter's 2-function metric depends on one radial function $\Delta(r)$ and one angular function $\Upsilon(\theta)$. 
Unfortunately the angular function $\Upsilon(\theta)$ will generically destroy the desired asymptotically Kerr behaviour,
so while the Carter 2-function metric is certainly mathematically interesting, it is not of direct astrophysical importance.

The  co-tetrad  for Carter's canonical 2-function spacetime is
\be \label{Carter-2-cotetrad}
\begin{split}
e^0=\sqrt{\frac{\Delta(r)}{r^2+a^2\cos^2(\theta)}}\;(dt-a\sin^2(\theta)d\phi); \quad 
e^1=\sqrt{\frac{r^2+a^2\cos^2(\theta)}{\Delta(r)}}\; dr;\\
e^2=\sqrt{\frac{r^2+a^2\cos^2(\theta)}{\Upsilon(\theta)}}\;d\theta; \quad 
e^3=\sqrt{\frac{\Upsilon(\theta)}{r^2+a^2\cos^2(\theta)}}\;\sin(\theta)\;(-a\,dt+(r^2+a^2)d\phi) .
\end{split}
\ee
The Kerr geometry corresponds to $\Delta(r)=r^2-2mr+a^2$ and $\Upsilon(\theta)=1$. 

However in general for the $
\theta\theta$ metric component we have
\be
g_{\theta\theta} = {\{r^2+a^2\cos^2(\theta)\}\over\Upsilon(\theta)}
\ee
To have an appropriate asymptotically flat limit $g_{\theta\theta}=r^2 + \O(1)$, or even an appropriate asymptotically Kerr limit  $g_{\theta\theta}=r^2+a^2\cos^2(\theta) + o(1)$, one is forced to take $\Upsilon(\theta)=1$.
Similar issues arise with the spacetime curvature. For instance the Ricci scalar is
\be
R = {\partial_r^2 \Delta(r) + \partial_\theta^2  \Upsilon(\theta)+3\cot(\theta) \partial_\theta  \Upsilon(\theta) -2  \Upsilon(\theta)
 \over r^2+a^2\cos^2(\theta)}
\ee
Trying to enforce suitable asymptotic falloff $o(r^{-2})$ for the Ricci scalar, given we already know  $\Delta(r)\sim r^2$,  again forces
 $\Upsilon(\theta)=1$.
Overall, we see that whereas the Carter 2-function metric is certainly of quite significant mathematical interest, it does not seem to be of direct astrophysical importance. 

\subsection{1-function version}

If we set  $\Upsilon(\theta)=1$ then the resulting 1-function Carter canonical  metric is a subset of the 3-function metrics considered above, corresponding to $\Phi(r)=0$ and $\Xi(r)=r$. 
Explicitly, the  co-tetrad  for Carter's canonical 1-function spacetime is
\be \label{Carter-1-cotetrad}
\begin{split}
e^0=\sqrt{\frac{\Delta(r)}{r^2+a^2\cos^2(\theta)}}\;(dt-a\sin^2(\theta)d\phi); \quad 
e^1=\sqrt{\frac{r^2+a^2\cos^2(\theta)}{\Delta(r)}}\; dr;\\
e^2=\sqrt{r^2+a^2\cos^2(\theta)}\;\;d\theta; \quad 
e^3={\frac{\sin(\theta)}{\sqrt{r^2+a^2\cos^2(\theta)}}}\;(-a\,dt+(r^2+a^2)d\phi) .
\end{split}
\ee
This restricted 1-function Carter canonical metric certainly enjoys many nice mathematical features, and is compatible with asymptotic flatness, but might reasonably be viewed as perhaps being a little too restrictive --- the 3-function metric that was the main theme of this article is somewhat more relaxed; that metric still satisfies the most central of the physically motivated restrictions, while being general enough to observationally interesting.

\section{Conclusions}

\enlargethispage{40pt}
From the discussion above we have learnt that the entire triply infinite class of three-function spacetimes parameterized by \{$\Delta(r)$, $\Phi(r)$, $\Xi(r)$\} and given by the following line element
\begin{equation}
\begin{split}
ds^2= & -\frac{\Delta(r)\exp(-2\Phi(r))-a^2\sin^2(\theta)}{\Xi(r)^2+a^2\cos^2(\theta)}dt^2
+\frac{\Xi(r)^2+a^2\cos^2(\theta)}{\Delta(r)}dr^2\\
&+(\Xi(r)^2+a^2\cos^2(\theta))d\theta^2
-2\; \frac{a(\Xi(r)^2-\Delta(r)\exp(-2\Phi(r))+a^2)\sin^2(\theta)}{\Xi(r)^2+a^2\cos^2(\theta)}dtd\phi\\
&+\frac{\left((\Xi(r)^2+a^2)^2-\exp(-2\Phi(r))\Delta(r)a^2\sin^2(\theta)\right)\sin^2(\theta)}{\Xi(r)^2+a^2\cos^2(\theta)}d\phi^2 .
\end{split}
\end{equation}
has a non--trivial Killing tensor given by equations (\ref{KT})--(\ref{KT2}) and (\ref{KT-tet}); implying the existence of a Carter constant.

Also, in these 3-function spacetimes one has the ``conservation law'' $\n_a[K,R]^a{}_b=0$. These conditions imply that both the Hamilton--Jacobi (geodesic) equation and the Klein--Gordon (wave) equation are separable in this entire class of  spacetimes. This class of spacetimes can also be arranged to be asymptotically Kerr by imposing
\be
\Delta(r)\sim r^2; \qquad \Phi=o\left(1\right); \qquad \Xi(r)\sim r.
\ee
Hence these spacetimes are particularly well-suited to modelling real astrophysical black holes. These spacetimes also admit Killing horizons which are located at the zeroes of $\Delta(r)$. That is, at  $\{r_H: \Delta(r_H)=0\}$. 
The angular velocities of these horizons are
\be
\Omega_{r_{H_i}}=\frac{a}{\Xi(r_{H_i}){}^2+a^2}.
\ee
This expression is independent of $\theta$, and hence the horizon rotates is if it were a solid body, thereby  yielding a ``rigidity'' result.
The area of the horizons are given by
\be
S(r_{H_i})=4\pi(\Xi(r_{H_i})^2+a^2),
\ee
and the surface gravity on the horizons is given by the remarkably simple and robust expression
\be
\kappa=\frac{\exp(-\Phi(r_{H_i}))\Delta'(r_{H_i})}{2(\Xi(r_{H_i})^2+a^2)} .
\ee
This expression is independent of $\theta$ and hence the horizons fulfil the zeroth law of black hole mechanics. 

\noindent The triply infinite class of 3-function spacetimes studied in this article exhibit an interesting (possibly optimal?) trade-off between generality and tractability. 
They cover a vast class of physically interesting spacetimes, as outlined in Table \ref{T:1} above, and also give toy models that can model the exterior spacetime of real astrophysical black holes. Such models will become particularly useful when probing the near-horizon strong-field dynamics of black holes, as we have already started doing with projects such as the event horizon telescope.

\section*{Acknowledgements}

JB was supported by a Victoria University of Wellington PhD Doctoral Scholarship.
\\
MV was directly supported by the Marsden Fund, 
via a grant administered by the Royal Society of New Zealand.


\end{document}